# On the Physical Meaning of Time-Domain Constitutive Models with Complex Parameters

## Nicos Makris

**Abstract** This paper revisits the physical meaning of linear, time-domain constitutive models with complex parameters that have been presented in the literature and concludes that such models are not physically realizable. While complex-parameter phenomenological models (including those with complex-order time derivatives) may be efficient in capturing in the frequency domain the frequency-dependent behavior of viscoelastic materials over a finite frequency band, they do not possess physically acceptable time-response functions. The paper first reviews the intimate relation between the causality of a physically realizable constitutive model and the analyticity of its frequency-response function and explains that in theory it is sufficient to conduct a nonlinear regression analysis for estimating the model parameters either on only the real part or on only the imaginary part of its frequency-response function, given that they are related with the Hilbert transform. Consequently, the resulting model-parameters are real-valued; therefore, there is no theoretical justification to conduct the nonlinear regression analysis for estimating the model parameters in the complex space. The paper concludes with an example by showing that the relaxation modulus of the complex-coefficient Maxwell model is a divergent function at all positive times; therefore it is not a physically realizable constitutive model.

## 1 Introduction

Traditional phenomenological models in mechanics and dynamics have real-valued parameters. In-series and parallel connections of frequency-independent springs with positive stiffnesses and dashpots with positive viscosities have been proposed to capture the frequency dependent behavior of viscoelastic materials and mechanical devices [1-4]. The stronger is the frequency-dependence of the linear mechanical behavior, the more are the springs and dashpots needed to be appended in the phenomenological network to satisfactorily capture the frequency-dependence of the mechanical behavior over a sufficiently large frequency range.

The time-dependent response of the material can then be given in terms of linear differential operators of integer order on the stress and strain, in which the coefficients are related to the moduli and viscosities of the components of the proposed network. More general linear differential operators on the stress and strain time-histories can be proposed by ignoring any candidate mechanical network with springs and dashpots,

Nicos Makris
Dept. of Civil and Environmental Engineering, Southern Methodist University, Dallas, Texas, USA
email: nmakris@smu.edu

Office of Theoretical and Applied Mechanics, Academy of Athens, 10679, Greece

and merely describe the measured linear viscoelastic response of the material with a phenomenological constitutive law of the form

$$\left[\sum_{m=0}^{M} a_m \frac{d^m}{dt^m}\right]\tau(t) = \left[\sum_{n=0}^{N} b_n \frac{d^n}{dt^n}\right]\gamma(t) \quad (1)$$

where, the coefficients $a_m$ and $b_n$ are restricted to real numbers and are the parameters of the constitutive model; whereas, the order of differentiation, $m$ and $n$ is restricted to integers. In this case, criteria must be applied to the coefficients $a_m$ and $b_n$ for the model to be physically acceptable [5-7]. The restrictions obtained on the frequency-independent coefficients $a_m$ and $b_n$ emerge from the requirement of a positive and monotonically decreasing relaxation function, which is sufficient to ensure that the work done by any imposed deformation starting from equilibrium is zero [6]. Furthermore, a necessary condition for any phenomenological constitutive model to be physically realizable is to be causal so that its response never precedes the excitation.

The linearity of equation (1) permits its transformation in the frequency domain

$$\tau(\omega) = [G_1(\omega) + iG_2(\omega)]\gamma(\omega) \quad (2)$$

where, $i = \sqrt{-1}$ = imaginary unit, $\tau(\omega) = \int_{-\infty}^{\infty}\tau(t)e^{-i\omega t}dt$ and $\gamma(\omega) = \int_{-\infty}^{\infty}\gamma(t)e^{-i\omega t}dt$ are the Fourier transforms of the stress and strain histories, and $\mathcal{G}(\omega) = G_1(\omega) + iG_2(\omega)$ is the dynamic modulus of the phenomenological constitutive model [1-3],

$$\mathcal{G}(\omega) = G_1(\omega) + iG_2(\omega) = \frac{\sum_{n=0}^{N} b_n(i\omega)^n}{\sum_{m=0}^{M} a_m(i\omega)^m} \quad (3)$$

The fidelity of equation (1) to capture experimentally measured data can be enhanced by either increasing the integer order of differentiation of the stress and strain differential operators (increasing $M$ and/ or $N$) or by generalizing the order of differentiation to real numbers (usually fractions). By introducing fractional-order time derivatives [8-14] the stress and strain differential operators appearing on the left and right-hand side of equation (1) become more efficient in capturing time-dependent viscoelastic behavior that follows power-laws; and therefore, a smaller number of parameters are needed in the proposed constitutive model [15-26 and references reported therein].

Regardless whether integer-order or fractional-order time derivatives are employed, the estimation of the parameters of any proposed viscoelastic model described by equation (1) is usually performed by best fitting the experimentally measured real and imaginary parts of the dynamic modulus, $\mathcal{G}(\omega) = G_1(\omega) + iG_2(\omega)$ given by equation (3). Accordingly, oscillatory tests are conducted over the frequency range that is relevant to the given application (finite frequency bandwidth) and the experimentally measured storage modulus, $G_1(\omega)$, and loss modulus, $G_2(\omega)$, are constructed over a sufficiently wide; yet, finite frequency range. A nonlinear regression numerical scheme is then employed to best estimate the parameters of the proposed candidate constitutive model ($a_m$, $b_n$ or even $m$ and $n$ in the event that a fractional derivative model is considered).

Given that in any experimental campaign the number of available experimental data is finite, a set of the model parameters that fits satisfactorily the real part, $G_1(\omega)$, usually underperforms when predicting the imaginary part, $G_2(\omega)$, and vice versa. In view of this challenge, Makris [23] and Makris and Constantinou [27] suggested to conduct the nonlinear regression for estimating the model parameters in the complex space. With this approach, the finite set of experimental data obtained for the storage modulus, $G_1(\omega)$, and loss modulus, $G_2(\omega)$, are analyzed as pairs of data (real and imaginary parts) corresponding to a single complex function, $\mathcal{G}(\omega) = G_1(\omega) + iG_2(\omega)$, rather than two independent sets of data belonging to two distinct real functions $G_1(\omega)$ and $G_2(\omega)$. In this way, the minimization of a complex objective function defined in a $L$-dimensional complex space, where $L$ is the entire number of the model parameters, inevitably yields complex-valued parameters [23,27].

When only the coefficients of the stress and strain differential operators in equation (1) are the model parameters ($L = 2 + M + N$); whereas the order of differentiation is restricted to integer numbers; the minimization of the complex objective function yields a phenomenological model with complex coefficients and integer-order time derivatives. When both the coefficients and the order of differentiation of the stress and strain differential operators in equation (1) are the model parameters $\left(L = 2(M + N + 1)\right)$ the minimization of the complex objective function yields a phenomenological model with complex coefficients and complex-order time derivatives [28-30]. Complex-parameter phenomenological models with complex-order time derivatives have been initially proposed to model the behavior of viscoelastic fluid dampers [23,27], and the dynamic response of foundations [31]. In electrical engineering and control theory complex-order time derivatives were proposed by Oustaloup *et al.* [32], while studies on the geometrical and physical meaning of complex-order derivative models have been presented by Nigmatullin and Le Mehaute [33]. Hartley *et al.* [34,35] introduced the concept of complex-conjugated order differentials which yield a real-valued output when subjected to a real-valued input. Subsequent studies on the implementation of complex-conjugated order derivatives in control theory have been presented by Adams *et al.* [36], Barbosa *et al.* [37], Machado [38] and Sathishkumar and Selvaganesan [39], among others. Complex-conjugated order derivatives, which is essentially a real differential operator, have been used by Atanackovic *et al.* [40] to study the response of a beam on a compliant foundation, while Atanackovic *et al.* [41] examined the constraints needed to be satisfied when complex-conjugated order derivatives are used in viscoelasticity.

Complex parameters in time-domain phenomenological models in mechanics have been introduced as early as 1940 in the seminal work of Theodorsen and Garrick [42] on the mechanism of flutter with the introduction of the frequency-independent structural damping which was later given the label "hysteretic damping" [43]. The appeal of frequency-independent dissipation generated by the "hysteretic dashpot" motivated a number of researchers to adopt a complex-valued stiffness in time-domain vibration analysis [44,45 among others], all being limited to harmonic steady-state excitations. However, Crandall [46,47] indicated the physical unrealizability of the "hysteretic dashpot" and its associated complex stiffness which violates causality.

Complex-parameter phenomenological models in the time-domain emerge merely because of the finite number of experimental data (finite frequency bandwidth) that are available when performing the nonlinear regression analysis for estimating the model-parameters. In reality, in any physically realizable constitutive model that satisfies causality, the real part, $G_1(\omega)$, and the imaginary part, $G_2(\omega)$, of its frequency-response function, $\mathcal{G}(\omega) = G_1(\omega) + iG_2(\omega)$ are intimately related with the Hilbert transform [48-51]. Consequently, in theory, assuming that a large number of experimental data is available over the entire frequency spectrum, a nonlinear regression analysis on only either the real-valued storage modulus, $G_1(\omega)$, or the real-valued loss modulus, $G_2(\omega)$, should be sufficient to estimate the real-valued parameters of a phenomenological constitutive model, given that $G_1(\omega)$ and $G_2(\omega)$ are Hilbert pairs.

The paper concludes that complex-parameter phenomenological models, including those with complex-order time derivatives are not physically realizable. They are merely practical engineering tools to approximate the dynamic behavior of viscoelastic materials and mechanical devices in the frequency domain that has been measured over a finite frequency range. Given that the real and imaginary parts of the frequency-response functions of a physically realizable material or device are Hilbert pairs and assuming that we know one of the two parts at all frequencies [52], it is sufficient to perform a regression analysis on a real objective function-either on the real part, $G_1(\omega) =$ storage modulus, or the imaginary part, $G_2(\omega) =$ loss modulus. Upon the real parameters of the phenomenological models are calibrated by fitting one of the components of the dynamic modulus $\left(\text{say the real part} = G_1(\omega)\right)$, the imaginary part, $G_2(\omega)$, is uniquely defined as the Hilbert transform of the real part, $G_1(\omega)$. Consequently, in theory, complex-parameter phenomenological models have no theoretical justification in classical mechanics and dynamics and are merely mathematical artifacts that do not have a time-domain representation in terms of a relaxation or a memory function.

## 2 Estimation of the parameters of a candidate constitutive model from oscillatory tests

The parameters of a proposed linear viscoelastic model can be identified either from oscillatory tests or time-dependent (creep or relaxation) measurements [1-3].

For instance, the dynamic modulus, $\mathcal{G}(\omega) = G_1(\omega) + iG_2(\omega)$ of a linear viscoelastic material can be measured experimentally by imposing a harmonic strain input of amplitude $\gamma_0$ and frequency, $\omega$ [1,2]

$$\gamma(t) = \gamma_0 \sin(\omega t) \tag{4}$$

and measuring the resulting stress, $\tau(t)$, that is needed to support the motion

$$\tau(t) = \tau_0 \sin(\omega t + \phi(\omega)) \tag{5}$$

where $\tau_0$ is the recorded stress amplitude and $\phi(\omega)$ is the frequency dependent phase lag of the stress to the strain history. Equation (5) can be expressed as

$$\tau(t) = G_1(\omega)\gamma_0 \sin(\omega t) + G(\omega)\gamma_0 \cos(\omega t) \tag{6}$$

where

$$G_1(\omega) = \frac{\tau_0}{\gamma_0} \cos\phi(\omega) \text{ and } G_2(\omega) = \frac{\tau_0}{\gamma_0} \sin\phi(\omega) \tag{7}$$

are the real (storage modulus) and imaginary (loss modulus) parts of the dynamic modulus of the viscoelastic material, $\mathcal{G}(\omega) = G_1(\omega) + iG_2(\omega)$. The energy dissipated per volume of material during each cycle of motion is

$$W_D(\omega) = \int_0^{\frac{2\pi}{\omega}} \tau(t) \frac{d\gamma(t)}{dt} \, dt = \pi \tau_0 \gamma_0 \sin\phi(\omega) \tag{8}$$

With the substitution of the result from equation (8) into the second equation in (7), the loss modulus of the material, $G_2(\omega)$ is expressed in terms of the measured energy dissipated per volume, per cycle, $W_D(\omega)$, (area of the stress-strain loop) and the imposed strain amplitude, $\gamma_0$

$$G_2(\omega) = \frac{W_D(\omega)}{\pi \gamma_0^2} \tag{9}$$

From equation (7), $G_1^2(\omega) + G_2^2(\omega) = \left(\frac{\tau_0}{\gamma_0}\right)^2$; therefore, the storage modulus of the viscoelastic material, $G_1(\omega)$ is given by

$$G_1(\omega) = \sqrt{\left(\frac{\tau_0}{\gamma_0}\right)^2 - G_2^2(\omega)} = \sqrt{\left(\frac{\tau_0}{\gamma_0}\right)^2 - \left(\frac{W_D(\omega)}{\pi \gamma_0^2}\right)^2} \tag{10}$$

Equation (10) indicates that the storage modulus, $G_1(\omega)$, of a linear viscoelastic material is directly related with its loss modulus, $G_2(\omega)$; therefore, it does not contain any additional information. Consequently, even without knowing that $G_1(\omega)$ and $G_2(\omega)$ are Hilbert pairs, assuming that we know $G_2(\omega)$ at all frequencies (therefore we know $G_1(\omega)$ from equation (10)), it is sufficient to perform a regression analysis for estimating the model parameters on the real-valued objective function, $G_2(\omega)$, in order to identify the real parameters of a candidate phenomenological constitutive model.

## 3  Causality and analyticity

The dynamic modulus, $\mathcal{G}(\omega) = G_1(\omega) + iG_2(\omega)$, given by equation (3) is a frequency-response function that relates a stress output to a strain input. When $m$ and $n$ are integers, the numerator on the right-hand side of equation (3) is a polynomial of degree $n$ and the denominator of degree $m$, therefore $\mathcal{G}(\omega)$ has $n$ zeros and $m$ poles. A frequency-response function that has more poles than zeros ($m > n$) is called strictly proper [53] and results in a strictly causal time response function, which means that it is zero at negative times and is finite at the time origin.

The stress, $\tau(t)$ in equation (1) can be computed in the time-domain with the convolution-integral

$$\tau(t) = \int_{-\infty}^{\infty} q(t - \xi)\gamma(\xi)d\xi \qquad (11)$$

where $q(t)$ is the memory function of the model [2,54,55], defined as the resulting stress at time $t$ due to an impulsive strain input at time $\xi$ ($\xi \leq t$) and is the inverse Fourier transform of the complex dynamic modulus,

$$q(t) = \frac{1}{2\pi}\int_{-\infty}^{\infty} \mathcal{G}(\omega)e^{i\omega t}d\omega \qquad (12)$$

The inverse Fourier transform given by equation (12) converges only when $\int_{-\infty}^{\infty}|\mathcal{G}(\omega)|d\omega < \infty$; therefore, $q(t)$ exists in the classical sense only when $\mathcal{G}(\omega)$ is a strictly proper transfer function ($m > n$). When the number of poles is equal to the number of zeros ($m = n$), the frequency-response function of the phenomenological model is simply proper and results to a time response function that has a weak singularity at the time origin because of the finite limiting value of the dynamic modulus at high frequencies. This means that, in addition to the hereditary effects, the model responds instantaneously to a given input. As an example, for the classical Maxwell model—that is a spring, $G = \frac{\eta}{\lambda}$, connected in series with a dashpot, $\eta$,

$$\tau(t) + \lambda\frac{d\tau(t)}{dt} = \eta\frac{d\gamma(t)}{dt} \qquad (13)$$

its memory function is [2,54,55]

$$q(t) = G\left[\delta(t-0) - \frac{1}{\lambda}e^{-\frac{t}{\lambda}}\right] \quad t \geq 0 \tag{14}$$

where $\delta(t-0)$ is the Dirac delta function at the time origin [56] and, $\lambda = \frac{\eta}{G}$, is the relaxation time of the Maxwell model.

When the dynamic modulus, $\mathcal{G}(\omega)$, of a phenomenological model is a simply proper transfer function ($m = n$), its complex viscosity, $\eta(\omega) = \frac{\mathcal{G}(\omega)}{i\omega}$, that is the transfer function from a strain-rate input to a stress output [1,2], is a strictly proper transfer function

$$\eta(\omega) = \eta_1(\omega) + i\eta_2(\omega) = \frac{\sum_{n=0}^{N} b_n (i\omega)^n}{\sum_{m=0}^{M} a_m (i\omega)^{m+1}} \tag{15}$$

The stress, $\tau(t)$, in equation (1) can be computed in the time domain with an alternative convolution integral

$$\tau(t) = \int_{-\infty}^{t} G(t-\tau) \frac{d\gamma(\tau)}{d\tau} d\tau \tag{16}$$

where $G(t)$ is the relaxation modulus of the viscoelastic material and is defined as the resulting stress at the present time, $t$, for a unit-step strain at time $\xi$ ($\xi \leq t$); and is the inverse Fourier transform of the complex viscosity [1,2]

$$G(t) = \frac{1}{2\pi} \int_{-\infty}^{\infty} \eta(\omega) e^{i\omega t} d\omega \tag{17}$$

Equation (17) holds only when the parameters $a_m$, $b_n$, $m$ and $n$ of the phenomenological constitutive model given by equation (1) are real-valued; therefore, it can be subjected to a real-valued unit-step excitation, $\gamma(t) = U(t-0)$, where $U(t-0)$ is the Heaviside unit-step function [56]. From equations (2) and (3), the frequency domain representation of the constitutive model given by equation (1) is:

$$\tau(\omega) = \frac{\sum_{n=0}^{N} b_n (i\omega)^n}{\sum_{m=0}^{M} a_m (i\omega)^m} \gamma(\omega) \tag{18}$$

The Fourier transform of a unit-step strain, $\gamma(t) = U(t-0)$, is $\gamma(\omega) = \pi\delta(\omega - 0) + \frac{1}{i\omega}$; therefore, for a unit-step loading, the real-parameter constitutive model given by equation (18) gives

$$\tau(\omega) = \frac{\sum_{n=0}^{N} b_n (i\omega)^n}{\sum_{m=0}^{M} a_m (i\omega)^m} \left[\pi\delta(\omega - 0) + \frac{1}{i\omega}\right] \tag{19}$$

and the relaxation modulus, $G(t)$, of the real-parameter constitutive model given by equation (1) is the inverse Fourier Transform of equation (19).

$$G(t) = \tau(t) = \frac{1}{2\pi} \int_{-\infty}^{\infty} \pi \delta(\omega - 0) \frac{\sum_{n=0}^{N} b_n (i\omega)^n}{\sum_{m=0}^{M} a_m (i\omega)^m} e^{i\omega t} d\omega$$
$$+ \frac{1}{2\pi} \int_{-\infty}^{\infty} \frac{\sum_{n=0}^{N} b_n (i\omega)^n}{\sum_{m=0}^{M} a_m (i\omega)^{m+1}} e^{i\omega t} d\omega \quad (20)$$

The fraction appearing in the second integral in the right-hand side of equation (20) is the complex dynamic viscosity, $\eta(\omega)$ given by equation (15); so equation (20) after taking $a_0 = 1$ assumes the form

$$G(t) = \tau(t) = \frac{1}{2} b_0 + \frac{1}{2\pi} \int_{-\infty}^{\infty} \eta(\omega) e^{i\omega t} d\omega \quad (21)$$

For fluid-like materials (infinite deformation under a constant load); $b_0 = 0$, and equation (21) reduces to equation (17).

For instance, for the fluid-like Maxwell model where a spring, $G = \frac{\eta}{\lambda}$, is connected in series with a dashpot, $\eta$, given by equation (13) ($b_0 = 0$), its strictly causal relaxation modulus is [2]

$$G(t) = G e^{-\frac{t}{\lambda}} \quad t \geq 0 \quad (22)$$

Note that the exponential decay on the right-hand side of equation (22) satisfies the requirement of a positive and monotonically decreasing relaxation modulus [5-7].

For the solid-like Kelvin-Voigt model that is a spring, $G$, connected in parallel with a dashpot, $\eta$, equation (1) reduces to

$$\tau(t) = G\gamma(t) + \eta \frac{d\gamma(t)}{dt} \quad (23)$$

For the Kelvin-Voight model given by equation (23), $b_0 = G$; whereas, its complex viscosity, $\eta(\omega)$, is

$$\eta(\omega) = \frac{G}{i\omega} + \eta \quad (24)$$

The Fourier transform of $1/(i\omega)$ is $\frac{1}{2}\operatorname{sgn} t$ (the signum function); whereas the Fourier transform of the constant, $\eta$, is $\eta \delta(t - 0)$. Accordingly, with $b_0 = G$, equation (21) results the following expression for the relaxation modulus of the Kelvin-Voight model.

$$G(t) = \frac{1}{2} G + \frac{1}{2} G \operatorname{sgn} t + \eta \delta(t - 0) = G U(t - 0) + \eta \delta(t - 0) \quad (25)$$

which is a causal time-response function with a weak singularity, $\eta \delta(t - 0)$, at the time origin due to the sudden stressing of the dashpot.

The Fourier transform relations between the relaxation modulus, $G(t)$, and the complex dynamic viscosity, $\eta(\omega)$, given by equation (17) for a fluid-like material or equation (21) for a solid-like material hold only when the parameters of the constitutive model given by equation (1) are real-valued; therefore, it can be subjected to a real-valued unit step-strain, $\gamma(t) = U(t-0)$. When the parameters of the constitutive model given by equation (1) are complex-valued the "unit-step strain" needs to be complex-valued — that is an analytic signal, as it is discussed in section 6 of this paper.

A necessary condition for any phenomenological constitutive model to be physically realizable is that its response never precedes the excitation. Accordingly, for a unit-step strain input, $\gamma(t) = U(t-0)$, where $U(t-0)$ is unit Heaviside step-function [50,56], the relaxation modulus $G(t)$ as defined by equation (17) has to be zero at negative times ($t < 0$) in order to satisfy causality. Using contour integration on the complex plane, the line integral appearing in equation (17) is expressed as

$$\int_{-\infty}^{\infty} \eta(\omega)e^{i\omega t}d\omega = \oint \eta(\omega)e^{i\omega t}d\omega - \int_{\Gamma} \eta(\omega)e^{i\omega t}d\omega = 0 \quad t < 0 \tag{26}$$

When time in equation (26) assumes negative values ($t < 0$), according to Jordan's lemma [57], the last integral in equation (26) vanishes if the arc $\Gamma$ with radius R ($\omega = R\,e^{i\theta}$) is in the bottom-half of the complex plane and the direction of integration is clockwise ($\sin(\theta) \leq 0$), as shown in Figure 1. Consequently, given a clockwise integration in the bottom-half of the complex plane, for a phenomenological model to be causal ($G(t) = 0$ for $t < 0$)

$$G(t) = \frac{1}{2\pi}\int_{-\infty}^{\infty} \eta(\omega)e^{i\omega t}d\omega = \frac{1}{2\pi}\oint \eta(\omega)e^{i\omega t}d\omega = 0 \quad t < 0 \tag{27}$$

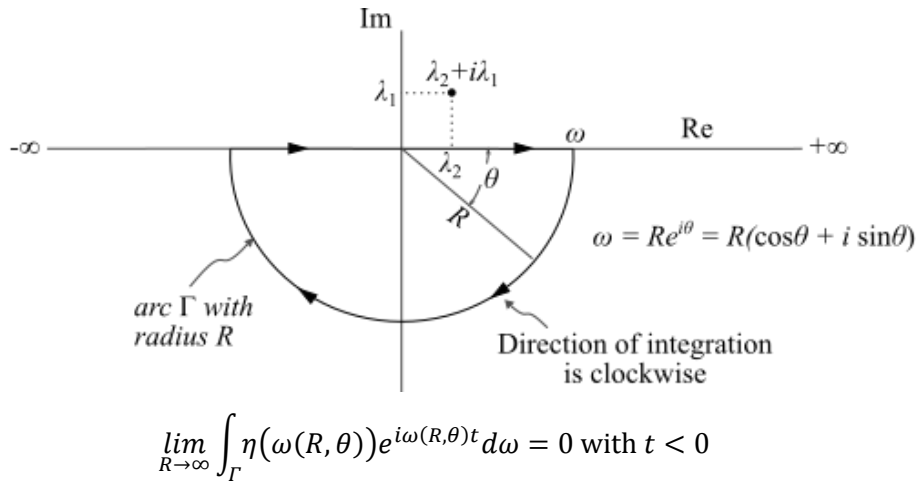

$$\lim_{R\to\infty}\int_{\Gamma} \eta(\omega(R,\theta))e^{i\omega(R,\theta)t}d\omega = 0 \text{ with } t < 0$$

**Figure 1**. For negative times, the integral $\int_{\Gamma}\eta(\omega)e^{i\omega t}d\omega$ vanishes as the integration is performed along the arc $\Gamma$ in the bottom-half complex plane ($\sin\theta \leq 0$) and the integration is performed clockwise.

its complex viscosity, $\eta(\omega)$, needs to be an analytic function in the bottom-half of the complex plane so that the contour (last) integral in equation (27) vanishes. This intimate relation between the causality of a time response function and the analyticity in the bottom-half of the complex plane of its corresponding frequency-response function emerges as the unique mathematical tool to examine the physical realizability of any proposed linear viscoelastic constitutive model.

## 4 Real and imaginary parts of a complex function that is analytic in the bottom half of the complex plane

Consider a complex function $f(z) = u(z) + iv(z)$ that is analytic in the bottom-half of the complex plane and let $\bar{\zeta} = \xi - i\eta$ be a point below the real axis, as shown in Figure 2. Clockwise integration of the function $f(z)/(z - \bar{\zeta})$ along the contour in the bottom-half complex plane gives

$$\oint \frac{f(z)}{z - \bar{\zeta}} dz = -2\pi i f(\bar{\zeta}) \tag{28}$$

Consider now the complex conjugate point of $\bar{\zeta}$, that is $\zeta = \xi + i\eta$ that is above the real axis; therefore, outside of the contour in the bottom-half complex plane. In this case

$$\oint \frac{f(z)}{z - \zeta} dz = 0 \tag{29}$$

Addition of equations (28) and (29) gives

$$\oint f(z) \left[ \frac{1}{z - \bar{\zeta}} + \frac{1}{z - \zeta} \right] dz = -2\pi i f(\bar{\zeta}) \tag{30}$$

Given that the contribution to the contour integral along the arc $\Gamma$ vanishes; we concentrate on the path along the real axis and we replace the variable $z$ with the real variable $x$. Accordingly, equation (30) gives

$$-f(\bar{\zeta}) = \frac{1}{2\pi i} \int_{-\infty}^{\infty} f(x) \left[ \frac{1}{(x - \xi) - i\eta} + \frac{1}{(x - \xi) + i\eta} \right] dx \tag{31}$$

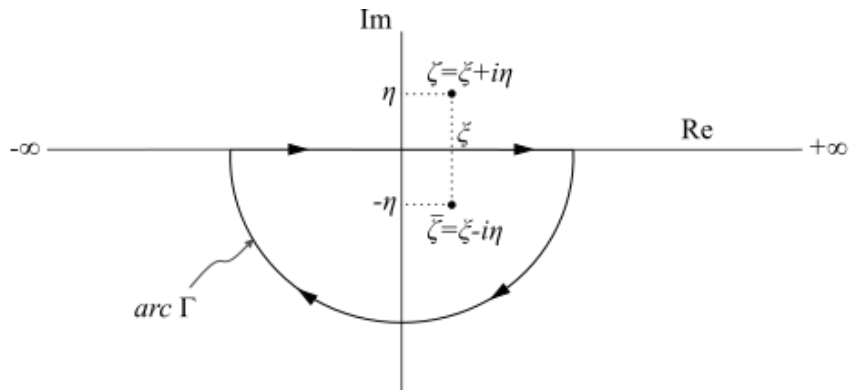

**Figure 2.** Clockwise integration in the bottom-half of the complex plane.

Using that $f(\bar{\zeta}) = u(\xi - i\eta) + iv(\xi - i\eta)$, and upon grouping the terms in the bracket of the integral, equation (31) gives

$$u(\xi,\eta) + iv(\xi,\eta) = \frac{i}{\pi} \int_{-\infty}^{\infty} [u(x,0) + iv(x,0)] \frac{x - \xi}{(x-\xi)^2 + \eta^2} dx \qquad (32)$$

Separation of the real and imaginary part of equation (32) gives

$$u(\xi,\eta) = -\frac{1}{\pi} \int_{-\infty}^{\infty} \frac{v(x-0)(x-\xi)}{(x-\xi)^2 + \eta^2} dx \qquad (33a)$$

$$v(\xi,\eta) = \frac{1}{\pi} \int_{-\infty}^{\infty} \frac{u(x-0)(x-\xi)}{(x-\xi)^2 + \eta^2} dx \qquad (33b)$$

For the limiting case where the pole $\bar{\zeta} = \xi - i\eta$ is on the real axis ($\eta \to 0$) equations (33a, b) reduce to

$$u(\xi) = -\frac{1}{\pi} \int_{-\infty}^{\infty} \frac{v(x)}{x - \xi} dx \qquad (34a)$$

$$v(\xi) = \frac{1}{\pi} \int_{-\infty}^{\infty} \frac{u(x)}{x - \xi} dx \qquad (34b)$$

The right-hand sides of equations (34a, b) are known as the Hilbert transform [49,51] and they show that the real and imaginary parts of an analytic function are Hilbert pairs. Note that the minus sign appearing in the right-hand side of equation (28) is because of the clockwise integration in the bottom-half of the complex plane, as shown in Figure 2, and it appears in the right-hand side of equation (34a). The divergence of the integrals appearing in equations (34a, b) when $x = \xi$ is allowed for by taking the Cauchy principal value of the integrals [48,49,51].

Equation (27) shows that for any causal, linear phenomenological constitutive model that its time response function is zero at negative times, its corresponding frequency-response function is analytic in the bottom-half complex plane. Furthermore, the foregoing analysis shows that when a complex function is analytic in the bottom-half of the complex plane, its real and imaginary parts are related with the Hilbert transform, as indicated by equations 34(a, b). Consequently in theory, assuming that a large number of experimental data is available over the entire frequency spectrum, it is sufficient to perform a nonlinear regression analysis for estimating the parameters of a candidate phenomenological constitutive model only on either the data of the real-valued loss modulus, $G_2(\omega) = \omega \eta_1(\omega)$, given by equation (9) or the data of the real-valued storage modulus, $G_1(\omega) = -\omega \eta_2(\omega)$, given by equation (10). Consequently, given that it is sufficient to perform the regression analysis for estimating the model parameters on a real function, the resulting parameters will be real-valued, and therefore, complex-parameter phenomenological models (including those with complex-order time derivatives) have no theoretical justification in classical mechanics. They are merely mathematical artifacts that can be viewed as practical engineering tools which are efficient in the frequency domain to approximate the dynamic behavior of linear viscoelastic materials and mechanical devices that has been measured over a finite frequency band.

## 5 Complex-order derivatives with conjugate orders

Non-integer differential calculus generalizes the classical differential calculus by introducing derivatives and integrals of real or even complex order [8-33]. The attraction of non-integer derivative constitutive models originates partly from their efficiency to capture the long-range fading memory of materials with a relatively small number of model parameters, and partly from their mathematical elegance.

Complex-order derivatives have been introduced to mathematics as early as 1930 by Post [28] and apparently first used to model viscoelastic behavior by Makris [23] and Makris and Constantinou [27]. Inevitably, complex-parameter constitutive models will result in a complex output even when subjected to a real-valued input. In an effort to bypass this impasse, Makris [58] examined the response of linear constitutive models with frequency-independent, complex coefficients and integer-order time derivatives, and by comparing their output with equivalent, higher-order constitutive models with frequency-dependent, real coefficients, the author concluded that the input in complex-parameter constitutive models needs to be complex-valued and more specifically, needs to be an analytic signal [51]. Analytic signals are complex-valued time-domain signals whose imaginary part is the Hilbert transform of their real part. While the use of analytic signals as input to complex-coefficient constitutive models allows for a mathematically consistent analysis in the frequency domain, the author at that time did not examine the physical meaning of the time-response functions (relaxation and memory functions) of complex-coefficient viscoelastic models. This task is accomplished in the last part of this paper.

Following the work of Makris [23] and Makris and Constantinou [27], complex-order derivatives of time-domain signals were proposed in control theory by Oustaloup *et al.* [32], where the output signal, $\tau(t)$, is the complex-order derivative of the input signal $\gamma(t)$.

$$\tau(t) = C \frac{d^{a+ib}}{dt^{a+ib}} \gamma(t) \tag{35}$$

In equation (35) $a$ and $b$ are real numbers, $i = \sqrt{-1}$ and $C$ is a complex-valued constant. In their paper, Oustaloup *et al.* [32] indicated very wisely that the action of the complex-order differential operator given by equation (35) is limited to some finite frequency bandwidth; however, they did not discuss what is the physical meaning of a complex output, $\tau(t)$, when the input signal, $\gamma(t)$, is a real-valued function. For instance, when the input signal, $\gamma = U(t-0)$, where $U(t-0)$ is the Heaviside unit-step function at the time origin; $\tau(t)$, is in theory the relaxation function of the network. Now given that the Laplace transform of $\gamma(t) = U(t - 0)$ is $1/s$, where $s$ is the Laplace variable, and that the Laplace transform of the complex derivative operator is [9]:

$$L\left\{\frac{d^{a+ib}}{dt^{a+ib}}\gamma(t)\right\} = s^{a+ib}\gamma(s) \tag{36}$$

the Laplace transform of equation (35) for a Heaviside unit step input $\gamma(t) = U(t - 0)$ is:

$$\tau(s) = C\frac{s^{a+ib}}{s} = Cs^{a-1+ib} = C\frac{1}{s^{1-a-ib}} \tag{37}$$

The relaxation function of the network described by equation (35) is the inverse Laplace transform of equation (37):

$$\tau(t) = G(t) = C\frac{1}{\Gamma(1 - a - ib)}\frac{1}{t^{a+ib}}U(t - 0) \tag{38}$$

with $1 - a > -1$ so $a < 2$ and $\Gamma(\cdot)$ being the Gamma function. Equation (38) shows that for the real-valued Heaviside unit-step input, $U(t - 0)$, the output, $\tau(t) = G(t)$ is complex-valued; and therefore, this result has a questionable physical meaning. To overcome this difficulty Hartley *et al.* [34,35] proposed the use of complex-conjugated order derivatives. By introducing the complex-conjugated order derivative, the differential operator appearing in equation (35) is augmented to

$$\tau(t) = B\left[\frac{d^{a+ib}}{dt^{a+ib}} + \frac{d^{a-ib}}{dt^{a-ib}}\right]\gamma(t) \tag{39}$$

where B is a constant.

When the input signal $\gamma(t) = U(t - 0)$ = Heaviside unit-step function at the time origin, the relaxation function of the network described by equation (39) is

$$\tau(t) = G(t) = B\left[\frac{1}{\Gamma(1 - a - ib)}\frac{1}{t^{a+ib}} + \frac{1}{\Gamma(1 - a + ib)}\frac{1}{t^{a-ib}}\right]U(t - 0) \tag{40}$$

By noting the property of the Gamma function: $\Gamma(\bar{z}) = \overline{\Gamma(z)}$, and after some algebraic manipulations, equation (40) yields [36]

$$\tau(t) = G(t) = \frac{2B}{\Gamma_1^2 + \Gamma_2^2}t^{-a}[\Gamma_1 \cos(b \ln t) + \Gamma_2 \sin(b \ln t)]U(t - 0) \tag{41}$$

where $\Gamma_1 = \Re\{\Gamma(1 - a + ib)\}$ and $\Gamma_2 = \Im\{\Gamma(1 - a + ib)\}$. When the constant coefficient, B, of the complex-conjugated order derivative model given by equation (39) is real-valued, its relaxation function given by equation (41) is also real-valued and the constitutive model given by (39) assumes physical meaning. The use of complex-conjugated order derivatives eliminates the imaginary contributions in the response; and therefore, the complex-conjugated order derivative operator appearing in equation (39) is a real operator. This has been pointed out by Adams *et al.* [36]; while Barbosa *et al.* [37] showed that when the complex-conjugated order derivative operator that appears in equation (39) operates on $\sin(\omega t)$, the result is real-valued. Consequently, when the coefficient B is real, the constitutive model expressed by equation (39) is a real-parameter model with clear physical meaning.

One possible attraction of the complex-conjugated order derivative operator, $\left[\frac{d^{a+ib}}{dt^{a+ib}} + \frac{d^{a-ib}}{dt^{a-ib}}\right]$ that is essentially a real operator, is that depending on the values of $a$ and $b$, it may result to a non-monotonically decreasing relaxation function due to the $\cos(b \ln t)$ and $\sin(b \ln t)$ appearing in the right-hand side of equation (41). The fluctuations of the resulting relaxation modulus at early times may be capable to model inertial effects in the material response [41]. However, inertial effects may also be captured effectively with second-order integer differential operators with real coefficients on the stress (inertoelastic and inertoviscoelastic models [59]).

## 6 The relaxation modulus of the complex-coefficient Maxwell model

While complex-parameter phenomenological constitutive models, in particular these with complex-order time derivatives, are efficient in approximating viscoelastic behavior over a sufficiently large; yet, finite frequency range, their physical non-realizability can be unveiled by examining their time response functions without invoking the argument of analyticity associated with a physically acceptable frequency-response function that leads to equations (34a) and (34b). The theoretical limitations of complex-parameter phenomenological models in the time domain are illustrated by examining the time response of the complex-coefficient Maxwell model with integer-order time derivatives:

$$\tau(t) + (\lambda_1 + i\lambda_2)\frac{d\tau(t)}{dt} = (\eta_1 + i\eta_2)\frac{d\gamma(t)}{dt} \tag{42}$$

where $\lambda_1 + i\lambda_2$ is a frequency-independent complex relaxation time ($\lambda_1 > 0$) and $\eta_1 + i\eta_2$ is a frequency-independent complex viscosity.

Phenomenological constitutive models with frequency-independent complex coefficients and integer-order derivatives such as the complex-coefficient Maxwell model given by equation (42) are equivalent to higher-order constitutive models with frequency-dependent real coefficients and integer-order derivatives [58]. For instance, the equivalent real-valued representation of the complex-coefficient Maxwell model given by equation (42) is:

$$\tau(t) + \lambda_1\frac{d\tau(t)}{dt} + \frac{\lambda_2}{|\omega|}\frac{d^2\tau(t)}{dt^2} = \eta_1\frac{d\gamma(t)}{dt} + \frac{\eta_2}{|\omega|}\frac{d^2\gamma(t)}{dt^2} \tag{43}$$

Recent progress in structural dynamics and the response modification of structures reveals that equation (43) essentially describes the behavior of a dashpot-inerter parallel connection that is connected in series with an elastic spring [59]. The term "inerter" has been introduced in control theory [60] as a two-node element in which the force (stress) output is proportional only to the relative acceleration (rate of the strain-rate) of its end nodes.

Given that real-valued constitutive models with frequency-dependent coefficients are equivalent to frequency-independent complex coefficient constitutive models; Makris [58] showed with frequency-domain analysis that frequency-independent complex-parameter models (such as the one given by equation (42)) yield precisely the same observable output (their real part) as the equivalent real-valued frequency-dependent models (such as the one given by equation (43)) only when the strain input is complex-valued and more specifically when the strain input is an analytic signal.

6.1  Analytic signals

When a complex-parameter phenomenological model (such as the one presented by equation (42)) is used, the induced excitation (strain-rate history = $\frac{d\gamma(\tau)}{d\tau}$, in the right-hand side of equation (42)) needs to be a complex-valued function [58]. Such complex signals in the time domain are known in optics and signal analysis as analytic signals [48,51]. For the real function $u(t)$ we may associate a complex function; $u(t) - iv(t)$ where $v(t)$ is the Hilbert transform of $u(t)$ as defined by equation (34b). In this case the complex function $u(t) - iv(t)$ is known as the analytic signal; and $v(t)$ is referred as the quadrature function of $u(t)$. For instance, the quadrature function of $\cos \omega t$ is $-\sin \omega t$ [49] and the analytic signal corresponding to $\cos \omega t$ is $\exp(i\omega t)$.

As an example, consider that the real-valued, frequency-dependent coefficient phenomenological model given by equation (43) is subjected to the harmonic input $\gamma(t) = \gamma_0 \cos \omega t$ where $\omega$ is the driving frequency. Given that the input excitation, $\gamma(t)$, contains a single frequency, $\omega > 0$, and that the constitutive model given by equation (43) is linear, the output stress-history assumes the expression, $\tau(t) = \tau_0 \cos(\omega t + \phi)$, where $\tau_0$ is a real-valued stress amplitude and $\phi$ is a real-valued phase difference. Substitution of the expressions, $\gamma(t) = \gamma_0 \cos \omega t$ and $\tau(t) = \tau_0 \cos(\omega t + \phi)$, in equation (43) gives

$$\tau_0[(1 - \omega\lambda_2)\cos(\omega t + \phi) - \omega\lambda_1 \sin(\omega t + \phi)] = -\gamma_0 \omega[\eta_1 \sin \omega t + \eta_2 \cos \omega t] \qquad (44)$$

We now apply the corresponding analytic signal of $\gamma_0 \cos \omega t$, that is $\gamma_0 \exp(i\omega t)$, to the frequency-independent, complex-coefficient Maxwell model given by equation (42) which is a linear model. Accordingly, the output stress-history assumes the expression $\tau(t) = \tau_0 \exp[i(\omega t + \phi)]$, where $\tau_0$ and $\phi$ are real-valued. Substitution of the expressions $\gamma(t) = \gamma_0 \exp(i\omega t)$ and $\tau(t) = \tau_0 \exp[i(\omega t + \phi)]$ in equation (42) gives:

$$\tau_0[1 + (i\omega\lambda_1 - \omega\lambda_2)][\cos(\omega t + \phi) + i\sin(\omega t + \phi)] \\ = \gamma_0(i\omega\eta_1 - \omega\eta_2)[\cos\omega t + i\sin\omega t] \tag{45}$$

Clearly, the imaginary part of the analytic strain input when multiplied with the imaginary terms of the complex-coefficient Maxwell model contributes to the real part of the response. After separating real and imaginary parts in equation (45), the reader concludes that the real part of equation (45), which is the observable quantity, is identical to equation (44), which is the outcome of the equivalent frequency-dependent, real-parameter model. More on the construction of the analytic signals of arbitrary recorded motions such as recorded earthquake accelerograms can be found in [58].

The concept of analytic signals can be extended to generalized functions [51]. For instance, the Hilbert transform of the Dirac delta function $\delta(t - 0)$ is evaluated with equation (34b) with the change of variables $\tau - t = \xi$; so that $d\tau = d\xi$:

$$v(t) = \frac{1}{\pi}\int_{-\infty}^{\infty} \frac{\delta(\tau - 0)}{\tau - t} d\tau = \frac{1}{\pi}\int_{-\infty}^{\infty} \frac{\delta(\xi + t)}{\xi} d\xi = -\frac{1}{\pi t} \tag{46}$$

and the analytic signal corresponding to $\delta(t - 0)$ is:

$$\theta(t - 0) = \delta(t - 0) + i\frac{1}{\pi t} \tag{47}$$

### 6.2 Evaluation of the relaxation modulus of the complex-coefficient Maxwell model

For the real-valued constitutive models, the relaxation modulus, $G(t)$, is defined as the resulting stress of the present time, $t$, for a unit-step strain $= U(t - 0)$ at the time origin; therefore the induced strain-rate is $\frac{d\gamma(t)}{dt} = \frac{dU(t-0)}{dt} = \delta(t - 0)$ [50,51]. In view of the foregoing analysis, the input excitation to the complex-parameter Maxwell model expressed by equation (42), has to be an analytic signal; therefore, the corresponding impulsive strain-rate that in theory will result to its relaxation modulus is the analytic signal given in equation (47). Accordingly, the relaxation modulus of the complex-parameter Maxwell model shall emerge as the solution of the following equation:

$$\tau(t) + (\lambda_1 + i\lambda_2)\frac{d\tau(t)}{dt} = (\eta_1 + i\eta_2)\left[\delta(t - 0) + i\frac{1}{\pi t}\right] \tag{48}$$

The Fourier transform of $\delta(t - 0)$ is unity; whereas the Fourier transform of $i/\pi t$ is $\text{sgn}(\omega)$, where $\text{sgn}(\omega)$ is the signum function. Accordingly, the Fourier transform of equation (48) gives:

$$\tau(\omega)[1 + i\omega(\lambda_1 + i\lambda_2)] = (\eta_1 + i\eta_2)[1 + \text{sgn}(\omega)] \tag{49}$$

and the inverse Fourier transform of equation (49) gives:

$$\tau(t) = G(t-0) = \frac{1}{2\pi}\int_{-\infty}^{\infty}\frac{\eta_1 + i\eta_2}{1 + i\omega(\lambda_1 + i\lambda_2)}e^{i\omega t}d\omega + \frac{1}{2\pi}\int_{-\infty}^{\infty}\frac{\eta_1 + i\eta_2}{1 + i\omega(\lambda_1 + i\lambda_2)}\text{sgn}(\omega)\,e^{i\omega t}d\omega \tag{50}$$

The fraction in each integral appearing in equation (50) is the complex viscosity (in the frequency domain) of the complex-parameter Maxwell model given by equation (42):

$$\eta(\omega) = \frac{\eta_1 + i\eta_2}{1 + i\omega(\lambda_1 + i\lambda_2)} \tag{51}$$

The denominator of equation (51) can be expressed as:

$$1 + i\omega(\lambda_1 + i\lambda_2) = i(\lambda_1 + i\lambda_2)\left(\omega - \frac{\lambda_2 + i\lambda_1}{\lambda^2}\right) \tag{52}$$

where $\lambda^2 = \lambda_1^2 + \lambda_2^2$. Given that, $\lambda_1 > 0$, the single pole of the complex viscosity, $\eta(\omega)$, given by equation (51), is above the real axis regardless what is the sign of $\lambda_2$ (see Figure 1). The first integral in the right-hand side of equation (50) is evaluated with the method of residues [48,57]

$$k(t) = \frac{1}{2\pi}\int_{-\infty}^{\infty}\frac{\eta_1 + i\eta_2}{1 + i\omega(\lambda_1 + i\lambda_2)}e^{i\omega t}d\omega = \frac{1}{2\pi}\int_{-\infty}^{\infty}\eta(\omega)\,e^{i\omega t}d\omega = \frac{\eta_1 + i\eta_2}{\lambda_1 + i\lambda_2}e^{i\frac{\lambda_2}{\lambda^2}t}e^{-\frac{\lambda_1}{\lambda^2}t} \tag{53}$$

At this point it is worth noting that since the single pole of the complex viscosity, $\eta(\omega)$, given by equation (51) is above the real axis; $\eta(\omega)$ is analytic in the bottom-half of the complex plane and therefore $k(t) = 0$ when time is negative ($t < 0$). The second integral in equation (50) is evaluated by recognizing that $\text{sgn}(\omega)$, is the Fourier transform of $i/\pi t$ and that the fraction expressed by equation (51) is the Fourier transform of the complex-valued function, $k(t)$, given by equation (53). Using the convolution theorem, the relaxation modulus of the complex-parameter Maxwell model given by equation (50) is expressed as:

$$G(t) = k(t) + \frac{i}{\pi}\int_{-\infty}^{\infty}k(\tau)\frac{1}{t - \tau}d\tau \tag{54}$$

which can be re-written as:

$$G(t) = k(t) - i\frac{1}{\pi}\int_{-\infty}^{\infty}\frac{k(\tau)}{\tau - t}d\tau \tag{55}$$

Our analysis concludes that the relaxation modulus, $G(t)$, of the complex-coefficient Maxwell model expressed by equation (55) is a complex-valued function that consists of the complex-valued function $k(t)$ offered by equation (53) minus $i$ times the Hilbert transform of $k(t)$. Traditionally in mathematics and signal processing the Hilbert transform is defined as the convolution of a real function, $u(x)$, of a real variable $x$ with the kernel $-1/\pi x$ as expressed by equation (34b) and presented in the literature [48,49,51]. Accordingly, the result reached by equation (55), which features the Hilbert transform of a complex-valued function, is a foreign concept in mathematical analysis and signal processing.

The puzzling result for the relaxation modulus, $G(t)$, of the complex-coefficient Maxwell model expressed by equation (55) pertains to any linear viscoelastic model expressed by equation (1) when the coefficients $a_m$ and $b_n$ are complex numbers. The linearity in equation (1) allows its transformation in the frequency domain:

$$\tau(\omega) = \eta(\omega)\dot{\gamma}(\omega) \tag{56}$$

where now $\dot{\gamma}(\omega) = \int_{-\infty}^{\infty} \dot{\gamma}(t)e^{-i\omega t}dt$ is the Fourier transform of the input strain-rate history and $\eta(\omega)$ is the complex viscosity of the complex-coefficient model given by equation (15).

When in equation (1) $m \geq n$, the complex viscosity, $\eta(\omega)$, given by equation (15) is a strictly proper transfer function [53] having $m + 1$ poles and $n < m + 1$ zeros. In this case its inverse Fourier transform

$$k(t) = \frac{1}{2\pi}\int_{-\infty}^{\infty} \eta(\omega)e^{i\omega t}d\omega \tag{57}$$

exists.

Now in the event that the coefficients $a_m$ and $b_n$ of the constitutive model described by equation (1) are complex, the input strain-rate history that will result to the relaxation modulus has to be the analytic signal $\dot{\gamma}(t) = \theta(t - 0)$ given by equation (47).

The Fourier transform of equation (47) is

$$\dot{\gamma}(\omega) = \int_{-\infty}^{\infty} \delta(t - 0)e^{-i\omega t}dt + \int_{-\infty}^{\infty} \frac{i}{\pi t}e^{-i\omega t}dt = 1 + \text{sgn}(\omega) \tag{58}$$

and equation (58) assumes the form

$$\tau(\omega) = \eta(\omega)(1 + \text{sgn}(\omega)) \tag{59}$$

The inverse Fourier transform of equation (59) is

$$\tau(t) = G(t) = \frac{1}{2\pi}\int_{-\infty}^{\infty} \eta(\omega)e^{i\omega t}d\omega + \frac{1}{2\pi}\int_{-\infty}^{\infty} \eta(\omega)\,\text{sgn}(\omega)\,e^{i\omega t}d\omega \tag{60}$$

The first integral in equation (60) is merely the complex function, $k(t)$, given by equation (57). The second integral in equation (60) is evaluated by recognizing that $\text{sgn}(\omega)$ is the Fourier transform of $i/\pi t$ and is essentially the convolution appearing the right-hand side of equation (54). Accordingly, the relaxation modulus, $G(t)$, of any viscoelastic model expressed by equation (1) when $a_m$ and $b_n$ are complex numbers and $m \geq n$, is expressed by equation (55), in which $k(t)$ is a complex function, rather than by equation (17) which holds only when $a_m$ and $b_n$ are real numbers.

Our analysis on evaluating the relaxation modulus, $G(t-0)$, of the complex-coefficient Maxwell model proceeds by ignoring the association of the second integral of equation (55) with the definition of the Hilbert transform and attempts to evaluate the integral:

$$I(t) = \frac{1}{\pi} \int_{-\infty}^{\infty} \frac{k(\tau)}{\tau - t} d\tau = \frac{\eta_1 + i\eta_2}{\lambda_1 + i\lambda_2} \frac{1}{\pi} \int_{-\infty}^{\infty} \frac{e^{i\frac{\lambda_2}{\lambda^2}\tau} e^{-\frac{\lambda_1}{\lambda^2}\tau}}{\tau - t} d\tau \tag{61}$$

Given that the complex viscosity of the complex-coefficient Maxwell model expressed by equation (61) is analytic in the bottom-half of the complex plane (see Figure 1), its Fourier transform, $k(t)$, as expressed by equation (53) is zero at negative times ($k(t) = 0$ for $t < 0$). Accordingly, the integral offered by equation (61) reduces to:

$$I(t) = \frac{1}{\pi} \int_{-\infty}^{\infty} \frac{k(\tau)}{\tau - t} d\tau = \frac{\eta_1 + i\eta_2}{\lambda_1 + i\lambda_2} \frac{1}{\pi} \int_0^{\infty} \frac{e^{\frac{i\lambda_2 - \lambda_1}{\lambda^2}\tau}}{\tau - t} d\tau \tag{62}$$

With the change of variables, $\tau - t = \xi$; $d\tau = d\xi$, equation (62) gives:

$$I(t) = \frac{1}{\pi} \frac{\eta_1 + i\eta_2}{\lambda_1 + i\lambda_2} e^{\frac{i\lambda_2 - \lambda_1}{\lambda^2}t} \int_{-t}^{\infty} \frac{e^{i\frac{\lambda_2}{\lambda^2}\xi} e^{-\frac{\lambda_1}{\lambda^2}\xi}}{\xi} d\xi \qquad t \geq 0 \tag{63}$$

From equation (53), the factor outside the integral of equation (63) is $k(t)/\pi$ and equation (63) assumes the form:

$$I(t) = \frac{1}{\pi} k(t) \left[ \int_{-t}^{0} \frac{e^{i\frac{\lambda_2}{\lambda^2}\xi} e^{-\frac{\lambda_1}{\lambda^2}\xi}}{\xi} d\xi + \int_0^{\infty} \frac{\cos\left(\frac{\lambda_2}{\lambda^2}\xi\right) e^{-\frac{\lambda_1}{\lambda^2}\xi}}{\xi} d\xi + i \int_0^{\infty} \frac{\sin\left(\frac{\lambda_2}{\lambda^2}\xi\right) e^{-\frac{\lambda_1}{\lambda^2}\xi}}{\xi} d\xi \right] \tag{64}$$

From Gradshteyn and Ryzhik ([61] page 497, section 3.941),

$$\int_0^{\infty} \cos\left(\frac{\lambda_2}{\lambda^2}\xi\right) \frac{e^{-\frac{\lambda_1}{\lambda^2}\xi}}{\xi} d\xi = \infty \quad \text{(diverges)} \tag{65a}$$

and

$$\int_0^{\infty} \sin\left(\frac{\lambda_2}{\lambda^2}\xi\right) \frac{e^{-\frac{\lambda_1}{\lambda^2}\xi}}{\xi} d\xi = \arctan\left(\frac{\lambda_2}{\lambda_1}\right), \text{ for } \lambda_1 > 0 \tag{65b}$$

With the results from equations (65a, b), equation (64) gives:

$$I(t) = \frac{1}{\pi} k(t) \left[ \int_{-t}^{0} \frac{e^{i\frac{\lambda_2}{\lambda^2}\xi} e^{-\frac{\lambda_1}{\lambda^2}\xi}}{\xi} d\xi + \infty + i \arctan\left(\frac{\lambda_2}{\lambda_1}\right) \right] \quad t \geq 0 \tag{66}$$

which is a divergent result at all times, including at $t = 0$.

Accordingly, the relaxation modulus of the complex-coefficient Maxwell model given by equation (55):

$$G(t) = k(t) - iI(t) \tag{67}$$

is a divergent function; therefore the complex-coefficient Maxwell model given by equation (42) does not have a physically realizable time-response function.

This section concludes that one of the simplest complex-parameter phenomenological models that is the complex-coefficient Maxwell model with integer-order time derivatives does not have a physically realizable relaxation modulus, therefore the complex-coefficient Maxwell model is not physically realizable.

## 7 Conclusions

This paper concludes that linear, time-domain constitutive models with complex parameters are not physically realizable. Upon reviewing the intimate relation between the causality of a physically realizable constitutive model and the analyticity of its corresponding frequency-response function in the bottom-half of the complex plane, the paper explains that in theory it is sufficient to conduct a nonlinear regression analysis for estimating the model-parameters of a candidate viscoelastic model either on only the real part or on only the imaginary part of its frequency-response function, given that the real and imaginary parts of a physically realizable frequency-response function are Hilbert pairs. Accordingly, assuming that we know one of the two parts at all frequencies, there is no theoretical justification to conduct the nonlinear regression analysis for estimating the model parameters in the complex space. The paper concludes that complex-parameter constitutive models, including those with complex-order time derivatives, are merely practical engineering tools to approximate in the frequency domain the dynamic behavior of viscoelastic materials which has been measured over a finite frequency range. However, these mathematical artifacts in the frequency domain do not possess a time-domain representation in terms of a relaxation or a memory function. This was illustrated by deriving the relaxation modulus of the complex-coefficient Maxwell model with integer-order time derivatives and it was found that the relaxation modulus is a divergent function at all times, including at the time origin. Consequently, the complex-coefficient Maxwell model is not physically realizable. Complex-conjugated order time derivatives is essentially a real differential operator since their imaginary contributions cancel; and when combined with real coefficients lead to a real-parameter constitutive model.

**Data Accessibility.** This work does not have any experimental data.

**Conflict of Interest.** The author does not have any conflict of interest.

**Funding.** This is not a funded research.**References**

1. Ferry, J.D.: Viscoelastic properties of polymers. New York, NY: J. Wiley and Sons (1980)
2. Bird, R.B., Armstrong, R.C., Hassager, O.: Dynamics of polymeric liquids. Vol 1: Fluid Mechanics, 2nd Edition. New York, NY: Wiley (1987)
3. Tschoegl, N.W.: The Phenomenological Theory of Linear Viscoelastic Behavior. An Introduction. Berlin, Heidelberg: Springer (1989)
4. Harris, C.M., Crede, C.E.: Shock and vibration handbook, 2nd Edition. New York, NY:McGraw-Hill (1976)
5. Brennan, M., Jones, R.S., Walters, K.: Linear viscoelasticity revisited. In Uhlherr PHT (ed) Proc X$^{th}$ Intern. Congr. Rheology, Vol 1. Sydney, Australia, Aug. 14-19. Australian Society of Rheology, 207-209 (1988)
6. Akyildiz, F., Jones, R.S., Walters, K.: On the spring-dashpot representation of linear viscoelastic behaviour. Rheologica. Acta. **29**, 482-484 (1990)
7. Beris, A.N., Edwards, B.J.: On the admissibility criteria for linear viscoelasticity kernels. Rheologica Acta **32**, 505-510 (1993)
8. Oldham, K.B., Spanier, J.: The Fractional Calculus. Mathematics in science and engineering Vol III, San Diego, CA: Academic Press Inc. (1974)
9. Miller, K.S., Ross, B.: An introduction to the fractional calculus and fractional differential equations. New York, NY: Wiley (1974)
10. Samko, S.G., Kilbas, A.A., Marichev, O.I.: Fractional Integrals and Derivatives: Theory and Applications. Amsterdam: Gordon and Breach Science (1974)
11. Carpinteri, A., Mainardi, F.: Fractals and fractional calculus in continuum mechanics. Springer (1997)
12. Podlubny, I.: Fractional Differential Equations. San Diego, CA: Academic Press (1999.
13. Mainardi, F.: Fractional Calculus and Waves in Linear Viscoelasticity: An Introduction to Mathematical Models. London, UK: Imperial College Press (2010)
14. Herrmann, R.: Fractional Calculus: An introduction for physicists. 2nd Edition, World Scientific Publishing Co. Pte. Ltd. (2014)
15. Gemant, A.: On fractional differentials. Philosophical Magazine **25**, 540-549 (2014)